\documentstyle[amssymb,preprint,aps]{revtex}

\begin{document}
\title{The phase diagram for the $(\lambda \phi ^{4}+\sigma \phi ^{2})_{2}$ model}
\author{E. Prodan}
\address{Rice University, Dept. of Physics-MS 61\\
6100 Main Street, Houston, TX 77005-1892}
\maketitle

\begin{abstract}
The two parameter model is reduced to a one parameter model by using simple
transformations. Because the separation between different phase regions for
a one parameter model is just a point, the equivalence between the two
models leads to the exact equation of the line that separates the broken and
un-broken phases in the $(\lambda ,\sigma )$ plane. Also, we obtain
nontrivial estimates on the stability region for this model.
\end{abstract}

\section{Introduction}

In this paper, we use the Euclidean strategy$^{1}$ combined with the
Gaussian approximation$^{2}$ to prove some exact results for the $\left(
\lambda \phi ^{4}+\sigma \phi ^{2}\right) _{2}$ model. The Euclidean
strategy can be summarized as it follows. First step is to consider a
(Euclidean) space cut-off interaction: 
\begin{equation}
U(\Lambda )=%
\textstyle\int%
_{\Lambda }d^{2}x:V(\phi (x)):_{m_{0}},
\end{equation}
where the normal ordering is with respect to the vacuum of the free field of
fixed mass $m_{0}$. The next step is the investigation of the cut-off
interacting measure: 
\begin{equation}
d\mu _{\Lambda }=Z^{-1}e^{-U\left( \Lambda \right)
+m_{0}^{2}/2\int_{R^{2}}d^{2}x:\left( \phi \left( x\right) -\xi \right)
^{2}:_{m_{0}}}d\mu _{m_{0},\xi }\text{,}
\end{equation}
where: 
\begin{equation}
Z=%
\textstyle\int%
e^{-U\left( \Lambda \right) +m_{0}^{2}/2\int_{R^{2}}d^{2}x:\left( \phi
\left( x\right) -\xi \right) ^{2}:_{m_{0}}}d\mu _{m_{0},\xi }.
\end{equation}
Along this paper, all the constants that normalize the measures to one will
be denoted by the same $Z$. In (2), $\mu _{m_{0},\xi }$ is the Gaussian
measure corresponding to the covariance $C_{m_{0}}=\left( -\Delta
+m_{0}^{2}\right) ^{-1}$ and mean $\xi $. Note that the additional term to $%
U\left( \Lambda \right) $ cancels the mass and the mean of the measure $\mu
_{m_{0},\xi }$ (see Ref. 3 for justification). The third step is the
investigation of the (thermodynamic) limit $\lim\nolimits_{\Lambda
\rightarrow R^{2}}\mu _{\Lambda }$. In the case when this limit is well
defined (and only in this case), one can construct the Minkowsky field by
using one of the well known reconstruction methods,$^{1}$ which represents
the last step of the Euclidean strategy. The cut-off theories can be defined
in more general settings, depending on what covariance one chooses to start
with or, more general, what boundary conditions are imposed. For example,
Ref. 1 analyses in detail the free and Dirichlet (also a half Dirichlet
boundary condition is considered) cut-off theories. Other boundary
conditions are discussed in Ref. 4, such as Neumann, periodic etc. These
boundary conditions are imposed directly on the covariance. In general, it
is not necessarily that the boundary conditions to be imposed on the
covariance. For example, adding a term like $%
\textstyle\int%
_{R^{2}\backslash \Lambda }d^{2}x:\delta V\left( \phi \left( x\right)
\right) :_{m_{0}}$ to the original interaction will also impose a boundary
condition. Such terms disappear in the thermodynamic limit and their
influence on the thermodynamic limit is trivial, unless we are situated in
the region of the dynamical instability. More exactly, suppose that the
potential $V\left( \phi \right) $ depends on some parameters. The region of
dynamical instability consists of those points in the parameters space for
which different boundary conditions leads to different thermodynamic limits.
In other words, a point is not in the dynamical instability region if the
states associated with various boundary conditions are identical and if this
state has a unique vacuum.$^{1}$ We call the separation between the stable
and unstable regions the manifold of dynamical instability. In general, a
quantum system consists of more than one pure phase (for definition see Ref.
5). If properly chosen, a boundary condition may help in isolating a pure
phase. For example, in Ref. 3, the vacuum sectors were found by expansions
around the mean field approximation which impose certain boundary
conditions. A thorough analysis of the $(\lambda \phi ^{4}+\sigma \phi
^{2})_{2}$ ($\lambda $%
\mbox{$<$}%
\mbox{$<$}%
$1)$ model$\prime $s vacuum and charged (soliton) sectors and how to
construct them by using different boundary conditions is given in Ref. 5. We
will show in the next sections that another way of generating useful
boundary conditions is the expansion around the Gaussian approximation.$^{2}$
The Gaussian approximation is defined by a certain choice of the covariance
plus the consideration of certain boundary conditions. As we will justify in
the following, by this procedure, one can always reduce by one the numbers
of parameters in a model. For $\left( \lambda \phi ^{4}+\sigma \phi
^{2}\right) $, this means that this model can be reduced to an equivalent
one parameter model. It is obvious that, for a one parameter model, the
instability manifold is given by discrete points. Strong evidence (also see
the conjecture propose in Ref. 1) shows that the instability manifold of the 
$\lambda \phi ^{4}+\sigma \phi ^{2}$ model is connected. This means that, in
the equivalent one parameter model, the instability manifold consists of one
and only one point. Using the relation between the equivalent models, one
can find the equation for the dynamical instability manifold in the $%
(\lambda $, $\sigma )$ plane. Moreover, one can also find nontrivial
estimates of the stability region for this model.

\section{Gaussian approximation for $(\protect\lambda \protect\phi ^{4}+%
\protect\sigma \protect\phi ^{2})_{2}$}

Let us fix the bare mass $m_{0}$ to an arbitrary value. Different
possibilities of choosing $m_{0}$ will be discussed later. The cut-off $%
(\lambda \phi ^{4}+\sigma \phi ^{2})_{2}$ model is defined by the
interacting measure: 
\begin{equation}
d\mu _{\Lambda }=Z^{-1}e^{-\int_{\Lambda }d^{2}x:\lambda \phi (x)^{4}+\sigma
\phi (x)^{2}:_{m_{0}}+m_{0}^{2}/2\int_{R^{2}}dx^{2}:\left( \phi \left(
x\right) -\xi \right) ^{2}:_{m_{0}}}d\mu _{m_{0},\xi },
\end{equation}
where the normal ordering is with respect to covariance $C_{m_{0}}=\left(
-\Delta +m_{0}^{2}\right) ^{-1}$. At this point, one can use the Gaussian
identity$^{4}$ to start with a different covariance. If we denote by $\mu
_{m,\xi }$ the Gaussian measure corresponding to the covariance $%
C_{m}=\left( -\Delta +m^{2}\right) ^{-1}$ and the same mean $\xi $, then: 
\begin{equation}
d\mu _{m_{0},\xi }=Z^{-1}e^{(m^{2}-m_{0}^{2})/2\int_{R^{2}}dx^{2}:\left(
\phi \left( x\right) -\xi \right) ^{2}:_{m_{0}}}d\mu _{m,\xi }.
\end{equation}
This provides the completely equivalent expression: 
\begin{equation}
d\mu _{\Lambda }=Z^{-1}e^{-\int_{\Lambda }d^{2}x:\lambda \phi (x)^{4}+\sigma
\phi (x)^{2}-\frac{m^{2}}{2}(\phi (x)-\xi )^{2}:_{m_{0}}+\delta V}d\mu
_{m,\xi },
\end{equation}
where: 
\begin{equation}
\delta V=m^{2}/2\int_{R^{2}\backslash \Lambda }dx^{2}:\left( \phi \left(
x\right) -\xi \right) ^{2}:_{m_{0}}.  \label{bc}
\end{equation}
We have isolated a term that can naturally define our boundary condition.
This boundary condition is given by the cancelation of $\delta V$ at the
exponent by adding the extra-term $\delta V$ to the original potential. One
can also change the normal ordering from the old to the new Gaussian
measure: 
\begin{equation}
:\phi \left( x\right) ^{n}:_{m_{0}}=\lim\limits_{a\rightarrow
0}\sum_{k=0}^{n}C_{n}^{k}\,\partial _{a}^{n-k}\exp [%
{\textstyle{a^{2} \over 8\pi }}%
\ln \frac{m_{0}^{2}}{m^{2}}+a\xi ]:\phi \left( x\right) ^{k}:_{\mu _{m,\xi
}},
\end{equation}
which finally leads to: 
\begin{equation}
d\mu _{\Lambda }=Z^{-1}e^{-\int_{\Lambda }d^{2}x:\lambda \phi
(x)^{4}+4\lambda \xi \phi (x)^{3}+\sigma ^{\prime }\phi (x)^{2}+\chi \phi
(x):_{\mu _{m,\xi }}}d\mu _{m,\xi },  \label{equivalent}
\end{equation}
with: 
\begin{eqnarray}
\sigma ^{\prime } &=&\sigma +\frac{3\lambda }{2\pi }(\ln 
{\textstyle{m_{0}^{2} \over m^{2}}}%
+4\pi \xi ^{2})-m^{2}/2,  \label{self-consistent} \\
\chi &=&\xi /\pi (3\lambda \ln 
{\textstyle{m_{0}^{2} \over m^{2}}}%
+4\pi \lambda \xi ^{2}+2\pi \sigma ).  \nonumber
\end{eqnarray}
The Gaussian approximation is defined by those values of $m$ and $\xi $ for
which $\sigma ^{\prime }$ and $\chi $ cancel. The expression (\ref
{equivalent}) of the interacting measure is equivalent to the original one
up to a boundary condition. The boundary condition is defined by the
equation (\ref{bc}). One can see now how the expansion around a Gaussian
point introduces the boundary condition: the particular values of $m$ and $%
\xi $ determine $\delta V$.

\section{Reduction to one parameter}

Let us make a few remarks about the meaning of the solutions of the
self-consistency equations (\ref{self-consistent}). Suppose these equations
have a solution, not necessarily unique. As we argued in the last section,
these solutions define different boundary conditions which leads to
different expression of the interacting measures. Even though these
expressions look very different, they differ by terms defined only on $%
R^{2}\backslash \Lambda $. Two questions must be answered. Do these
interacting measures have a thermodynamic limit? If the answer is yes, then
we found the vacuum states. The energy per unit length of these states is
the same because the contributions from the boundary conditions cancels in
the thermodynamic limit. And here comes the second question. If the
thermodynamic limits exist, are they really different? If the answer is yes,
than the ground state is degenerate and, possibly, a symmetry breaking takes
place.

We start now analyzing the solutions of the self-consistency equations (10).
We will show that, in some limiting cases, one can prove the thermodynamic
limit of the interacting measures corresponding to some of the solutions of
the self-consistency equations. Moreover, in these limiting cases, one can
even answer the second question. There will be always the trivial solution: 
\begin{equation}
\begin{array}{l}
\xi =0 \\ 
m^{2}=\frac{3\lambda }{\pi }W_{0}(%
{\textstyle{\pi m_{0}^{2} \over 3\lambda }}%
\exp 
{\textstyle{2\pi \sigma  \over 3\lambda }}%
),
\end{array}
\label{first-solution}
\end{equation}
where $W_{0}$ is the Lambert function of rank zero. The interacting measure
corresponding to this solution is written in Eq. (\ref{lambdazero}). This
equivalence has been actually found long time ago.$^{8}$ Based on this
observation, Guerra, Rosen and Simon concluded in Ref. 9 that $\left(
\lambda \phi ^{4}+\sigma \phi ^{2}\right) _{m_{0}}$ has as many echilibrium
states as $\left( \lambda \phi ^{4}\right) _{m}$ has. The other solutions
must satisfy: 
\begin{equation}
\begin{array}{l}
8\lambda \xi ^{2}=m_{0}^{2}\exp 
{\textstyle{2\pi  \over 3\lambda }}%
\left( 2\lambda \xi ^{2}+\sigma \right) \\ 
m^{2}=8\lambda \xi ^{2}.
\end{array}
\label{system}
\end{equation}
In this case, there are two independent solutions: 
\begin{equation}
\begin{array}{l}
\xi ^{2}=%
{\textstyle{-3 \over 4\pi }}%
W_{0,-1}(-%
{\textstyle{\pi m_{0}^{2} \over 6\lambda }}%
\exp 
{\textstyle{2\pi \sigma  \over 3\lambda }}%
) \\ 
m^{2}=8\lambda \xi ^{2},
\end{array}
\label{second-solution}
\end{equation}
where $W_{0,-1}$ are the Lambert functions of rank $0$ and $-1$. These
solutions become very useful in certain limits of the coupling constants,
when the Gaussian approximation is very precise. Let us discuss first $%
\sigma >0$ case, when the potential $\lambda \phi ^{4}+\sigma \phi ^{2}$ has
only one minimum at $\phi =0$. In this case, one may expect the symmetry $%
\phi \rightarrow -\phi $ to be unbroken. However, this is happening only for 
$\lambda $ small. Using solution (\ref{first-solution}), the expression (\ref
{equivalent}) of the interacting measure become: 
\begin{equation}
d\mu _{\Lambda }=Z^{-1}e^{-\lambda \int_{\Lambda }dx^{2}:\phi (x)^{4}:_{\mu
_{m\xi =0}}}d\mu _{m,\xi =0}\text{.}  \label{lambdazero}
\end{equation}
One can use the scalling identity$^{3}$ to normalize the mass to unity: 
\begin{equation}
d\mu _{\Lambda }=Z^{-1}e^{-\frac{\lambda }{m^{2}}\int_{\Lambda ^{\prime
}}dx^{2}:\phi (x)^{4}:_{\mu _{1\xi =0}}}d\mu _{1,\xi =0}\text{,}
\label{lamdazeroscaled}
\end{equation}
where $\Lambda ^{\prime }=\Lambda /m^{2}$. The limit $\lambda \rightarrow 0$
does not automatically bring us in the small coupling regime, because, as
one can see from (\ref{lamdazeroscaled}), the small coupling regime is
defined by $\lambda /m^{2}\rightarrow 0$. However, $m\rightarrow 2\sigma $
as $\lambda $ goes to zero so, indeed, $\lambda \rightarrow 0$ defines the
small coupling regime. Then the thermodynamic limit of (\ref{lambdazero})
can be achieved by an ordinary cluster expansion,$^{6}$ and it is already
known that $\langle \phi \rangle =0$ in the thermodynamic limit. Moreover,
the system has only one pure phase in $\lambda \rightarrow 0$ limit. For
large $\lambda $, the cluster expansion will no longer work for expression (%
\ref{lambdazero}), so we need to find a better approximation to start with.
Guided by the conjecture proposed in Ref. 1, we guess that the symmetry $%
\phi \rightarrow -\phi $ is broken in the $\lambda \rightarrow \infty $
limit and we try a solution with $\xi \neq 0$. The right one is: 
\begin{equation}
\begin{array}{l}
\xi ^{2}=%
{\textstyle{-3 \over 4\pi }}%
W_{-1}(-%
{\textstyle{\pi m_{0}^{2} \over 6\lambda }}%
\exp 
{\textstyle{2\pi \sigma  \over 3\lambda }}%
) \\ 
m^{2}=8\lambda \xi ^{2},
\end{array}
\label{secsol}
\end{equation}
because $m^{2}$ and $\xi ^{2}$ go to infinity as $\lambda \rightarrow \infty 
$. As we will argue later, this is enough for the convergence of the cluster
expansion around this solution and, thus, the thermodynamic limit exists.
One can ask what is happening with the solution $\xi =0$ in this limit. For $%
\lambda \rightarrow \infty $, the thermodynamic limit of (\ref{lambdazero})
cannot be proven with the existing techniques. Supposing the thermodynamic
limit exists and can be found by some other methods, then more likely the
thermodynamic limit of (\ref{lambdazero}) is not ergodic but it can be
decomposed in the two broken phases, corresponding to the solutions (\ref
{secsol}). The last thing we want to mention for $\sigma >0$ case is the
semi-classical mass. This is defined by: 
\begin{equation}
m_{c}^{2}=V^{\prime \prime }(\phi )|_{\phi =0}=2\sigma \text{.}
\end{equation}
As we already mentioned, in the $\lambda \rightarrow 0$ limit, $m$ goes to
this semiclassical value, independent of the value of $m_{0}$. Moreover, if
one starts with $m_{0}=m_{c}$, then the solution given by Eq. (\ref
{first-solution}) reduces to $\xi =0$ and $m=m_{c}$. In other words, this
particular solution of the self-consistency equations coincides with the
semiclassical approximation.

If $\sigma <0$, the potential $\lambda \phi ^{4}+\sigma \phi ^{2}$ has two
minima located at: 
\begin{equation}
\phi =\xi _{c}=\pm \sqrt{-\sigma /(2\lambda )}\text{.}
\end{equation}
In this case, one may expect that the symmetry $\phi \rightarrow -\phi $ to
be broken. This is true at least for $\lambda \rightarrow 0$ and $\lambda
\rightarrow \infty $ limits and, as we shall see, it is not true for
intermediate values. For $\lambda \rightarrow 0$, the useful solution is: 
\begin{equation}
\begin{array}{l}
\xi ^{2}=%
{\textstyle{-3 \over 4\pi }}%
W_{0}(-%
{\textstyle{\pi m_{0}^{2} \over 6\lambda }}%
\exp 
{\textstyle{2\pi \sigma  \over 3\lambda }}%
) \\ 
m^{2}=8\lambda \xi ^{2},
\end{array}
\end{equation}
because $\xi ^{2}\rightarrow \infty $ and $m$ goes to the finite value $%
-4\sigma $ as $\lambda \rightarrow 0$. For $\lambda \rightarrow \infty $,
the useful solution is: 
\begin{equation}
\begin{array}{l}
\xi ^{2}=%
{\textstyle{-3 \over 4\pi }}%
W_{-1}(-%
{\textstyle{\pi m_{0}^{2} \over 6\lambda }}%
\exp 
{\textstyle{2\pi \sigma  \over 3\lambda }}%
) \\ 
m^{2}=8\lambda \xi ^{2},
\end{array}
\end{equation}
because $\xi ^{2}\rightarrow \infty $ and the mass increases even faster as $%
\lambda \rightarrow \infty $. We remember that, if $\xi $ and $m$ are
solutions of the self-consistency equation, then expression (\ref{equivalent}%
) of the interacting measure reduces to: 
\begin{equation}
d\mu _{\Lambda }=Z^{-1}e^{-\int_{\Lambda }dx^{2}:\lambda \phi
(x)^{4}+4\lambda \xi \phi (x)^{3}:_{\mu _{m,\xi }}}d\mu _{m,\xi }\text{.}
\end{equation}
For all $\xi \neq 0$ solutions, $m=8\lambda \xi ^{2}$. By using the scale
transformation,$^{3}$ we can normalize the mass to one in the above
expression: 
\begin{equation}
d\mu _{\Lambda }=Z^{-1}e^{-\int_{\Lambda ^{\prime }}dx^{2}:\frac{1}{8\xi ^{2}%
}\phi (x)^{4}+\frac{1}{2\xi }\phi (x)^{3}:_{\mu _{1,\xi }}}d\mu _{1,\xi }%
\text{,}  \label{reduced}
\end{equation}
where $\Lambda ^{\prime }=\Lambda /m^{2}$. One can see now that the small
coupling regime is achieved when $\xi \rightarrow \infty $ and all the cases
when this is happening have been discussed above. However, even for large
values of $\xi $, the thermodynamic limit of (\ref{reduced}) cannot be
achieved by an ordinary cluster expansion. This is due to the fact that the
potential $\frac{1}{8\xi ^{2}}\phi ^{4}+\frac{1}{2\xi }\phi ^{3}$ is not
uniformly bounded from below as $\xi \rightarrow \infty $. When the field is
localized near $-\xi $, the potential behaves as $-\xi ^{2}$. Nevertheless,
an expansion is phase boundaries$^{3}$ or a cluster expansion with
small/large field conditions$^{7}$ converges in the thermodynamic limit. The
result is:$^{2}$%
\begin{equation}
\langle \phi (x)\rangle _{\Lambda \rightarrow R^{2}}=\xi +o(\xi ^{-1}),
\end{equation}
which shows that the symmetry $\phi \rightarrow -\phi $ is broken in the
limit $\xi \rightarrow \infty $. For $\sigma <0$ case, one can also define a
semiclassical mass: 
\[
m_{c}^{2}=V^{\prime \prime }(\phi )|_{\phi =\xi _{c}}=-4\sigma \text{.} 
\]
We already mentioned that $m\rightarrow m_{c}$ as $\lambda \rightarrow 0$.
Moreover, if one starts with $m_{0}=m_{c}$, then $\xi =\xi _{c}$ and $%
m=m_{c} $ is a solution of the self-consistency equations. In other words,
this particular solution coincides with the semiclassical approximation.

Yet we discussed only how the Gaussian approximation can be combined with
the standard results of the cluster expansion technique to analyze the $%
(\lambda \phi ^{4}+\sigma \phi ^{2})_{2}$ model in certain limits of the
coupling constants. However, the most important result of this section are
the equivalent expressions (\ref{lamdazeroscaled}) and (\ref{reduced}) of
the interacting measure. This expressions characterize the system for all
allowed values of the coupling constants, because the solutions of the
self-consistancy equations that leaded to these exressions exist for all
allowed values of the coupling constants. These equivalent forms of the
interacting measure depend only on one parameter because the dependency of $%
\Lambda ^{\prime }$ on $m$ is irrelevant in the thermodynamic limit.

\section{The instability manifold}

We remind that the expressions (\ref{lamdazeroscaled}) and (\ref{reduced})
are equivalent to the original interacting measure up to a boundary
condition. The difference between them is the extra-term given by equation (%
\ref{bc}). This means that we should be able to reconstruct the phase
diagram of the original model by imposing different boundary conditions on
these expressions. Let us consider the first one: 
\begin{equation}
d\mu _{\Lambda }=Z^{-1}e^{-\tilde{\lambda}\int_{\Lambda ^{\prime
}}dx^{2}:\phi (x)^{4}:_{\mu _{1\xi =0}}}d\mu _{1,\xi =0}\text{,}
\label{firstexpr}
\end{equation}
where $\tilde{\lambda}=\lambda /m^{2}$. As we argued in the previous
section, the symmetry is unbroken for $\tilde{\lambda}$ small and it is
broken for $\tilde{\lambda}$ large. This means that the instability manifold
has at least one point. Assuming that the instability manifold for $\left(
\lambda \phi ^{4}+\sigma \phi ^{2}\right) $ is connected, then the
instability manifold of (\ref{firstexpr}) must contain one and only one
point, denoted by $\tilde{\lambda}_{cr}$ in the following. This is an
absolute constant (a number). Using the exression of $m$ given by the
solution (\ref{first-solution}), we can find the equation of the instability
manifold in the $\left( \lambda ,\sigma \right) $ plane: 
\begin{equation}
\frac{\lambda }{\frac{3\lambda }{\pi }W_{0}(%
{\textstyle{\pi m_{0}^{2} \over 3\lambda }}%
\exp 
{\textstyle{2\pi \sigma  \over 3\lambda }}%
)}=\tilde{\lambda}_{cr}\text{,}  \label{lacritic}
\end{equation}
or equivalently: 
\begin{equation}
\sigma =\frac{3\lambda }{2\pi }\ln [c_{cr}\lambda /m_{0}^{2}]\text{,}
\label{criticalEq}
\end{equation}
where $c_{cr}$ is an absolute constant. This is the exact analytic equation
of the instability manifold. The second exression is even more useful
because it allows an evaluation of $\tilde{\lambda}_{cr}$. Because 
\begin{equation}
d\mu _{+}=Z^{-1}e^{-\int_{x\in \Lambda ^{\prime }}:%
{\textstyle{1 \over 8\xi ^{2}}}%
\phi \left( x\right) ^{4}+%
{\textstyle{1 \over 2\xi }}%
\phi \left( x\right) ^{3}:_{1,\xi }}d\mu _{C_{1},\xi }  \label{miu+}
\end{equation}
and: 
\begin{equation}
d\mu _{-}=Z^{-1}e^{-\int_{x\in \Lambda ^{\prime }}:%
{\textstyle{1 \over 8\xi ^{2}}}%
\phi \left( x\right) ^{4}-%
{\textstyle{1 \over 2\xi }}%
\phi \left( x\right) ^{3}:_{1,-\xi }}d\mu _{C_{1},-\xi }  \label{miu-}
\end{equation}
differ by just a boundary condition, it is sufficient to consider only $\xi
\in (0,\infty )$. For the one parameter model, the instability manifold is
given by discrete points. As we already discussed, in the limit $\xi
\rightarrow \infty $, the system is composed from more than one pure phase,
the two measures (\ref{miu+}) and (\ref{miu-}) leads to different states in
the thermodynamic limit. This means that, for $\xi $ large we are in the
region of dynamical instability. Also, we know that for $\lambda \rightarrow
0$ and $\sigma >0$ ($\xi \rightarrow 0$) the system is composed from a
single pure phase. In this limit, the quantum fluctuations are so large that
the symmetry $\phi \rightarrow -\phi $ is restored and the two measures, (%
\ref{miu+}) and (\ref{miu-}), lead to the same state in the thermodynamic
limit. This means that, for $\xi \rightarrow 0$, we are in the region of
dynamical stability. In conclusion, the manifold of dynamical instability
for the one parameter model has at least one point. If we assume that the
dynamical instability manifold for $(\lambda \phi ^{4}+\sigma \phi ^{2})_{2}$
model is connected, then the instability manifold for the one parameter
model must contain one and only one point (different points will lead to
un-connected lines in the ($\lambda ,\sigma $) plane). We denote by $\xi
_{cr}$ this critical point, which is an absolute constant (a number). We
remind that all different solutions with $\xi \neq 0$ that we found above
came from the same system of equations (\ref{system}). In consequence, the
manifold of dynamical instability in $(\lambda ,\sigma )$ plane is given by
the following equation: 
\begin{equation}
8\lambda \xi _{cr}^{2}=m_{0}^{2}\exp [%
{\textstyle{4\pi  \over 3}}%
\xi _{cr}^{2}+\frac{2\pi }{3\lambda }\sigma ],
\end{equation}
or equivalently: 
\begin{equation}
8\xi _{cr}^{2}\exp [-%
{\textstyle{4\pi  \over 3}}%
\xi _{cr}^{2}]\lambda =m_{0}^{2}\exp [\frac{2\pi }{3}\sigma /\lambda ].
\end{equation}
Denoting 
\begin{equation}
c_{cr}\equiv 8\xi _{cr}^{2}\exp [-%
{\textstyle{4\pi  \over 3}}%
\xi _{cr}^{2}]\text{,}  \label{ccritic}
\end{equation}
which is again an absolute constant, we obtain the same equation (\ref
{criticalEq}) for the instability manifold. For practical purposes, it is
important to give an estimate of $c_{cr}$. One possible way to evaluate this
constant is to explore the region of dynamical stability by cluster
expansion. Unfortunately, as it follows from Ref. 6, this technique does not
allow yet very sharp estimates. However, we can give an upper estimate of $%
c_{cr}$ from its analytic expression (\ref{ccritic}): 
\begin{equation}
c_{cr}\leqslant \sup_{x\in R_{+}}8x\exp [-%
{\textstyle{4\pi  \over 3}}%
x]=6/(e\pi )\approx 0.70\text{.}
\end{equation}
Let us consider the family of curves in the $(\lambda ,\sigma )$ plane given
by: 
\begin{equation}
\sigma =\frac{3\lambda }{2\pi }\ln [c\lambda /m_{0}^{2}]\text{ \ \ , }c\in
(0,\infty )\text{.}
\end{equation}
For $c>c_{cr}$, these curves lie on the stability region while for $c<c_{cr}$
they lie on the instability region. Then, because $c_{cr}\leqslant 6/(e\pi )$%
, the region spanned by the curves with $c>6/(e\pi )$ belongs to the
stability region. We can also give a simple method for checking if a point
in the $(\lambda ,\sigma )$ plane belongs to the stability region. For a
given $\lambda $ and $\sigma $, if: 
\begin{equation}
(\lambda /m_{0}^{2})^{-1}e^{\frac{2\pi \sigma }{3\lambda }}>6/(e\pi )\text{,}
\end{equation}
then the point is situated in the region of dynamical stability. For
example, if $\sigma =0$, the symmetry is not broken at least for $\lambda
/m_{0}^{2}<e\pi /6\approx 1.42$. Another important conclusion of the above
analysis is that the $(\lambda \phi ^{4}+\sigma \phi ^{2})_{2}$ model is
completely determined by $\xi $. In the limit $\xi \rightarrow \infty $,
this parameter is approximately given by: $\xi \approx \langle \phi \rangle $
which is a physically measurable quantity.

\bigskip

{\bf References\medskip }\newline
$^{1}$B. Simon, {\it The\ }$P\left( \phi \right) _{2}${\it \ Euclidean
(Quantum) Field Theory} (Princeton Univ. Press, Princeton, 1974).\newline
$^{2}$E. Prodan, quant-ph/9905088 (1999).\newline
$^{3}$J. Glimm, A. Jaffe and T. Spencer, Ann. Phys. {\bf 101}, 610 (1976).%
\newline
$^{4}$J. Glimm, A. Jaffe, {\it Quantum Physics} (Springer, Berlin, 1981).%
\newline
$^{5}$J. Fr\"{o}hlich, in {\it Invariant Wave Equation,} Proc. of Eltore
Majorana (1978).\newline
$^{6}$E. Prodan, J. Math. Phys. {\bf 41},787 (2000).\newline
$^{7}$V. Rivasseau, CMR Proceedings and Lecture Notes {\bf 7}, 99 (1994).%
\newline
$^{8}$R. Baumel, Princeton University Thesis (1973).\newline
$^{9}$F. Guerra, L. Rosen and B. Simon, Ann. Math. {\bf 101}, 111 (1975).

\end{document}